\documentclass[aps,prb,reprint,superscriptaddress]{revtex4-1}

\pdfoutput=1  

\usepackage{amssymb}
\usepackage{graphicx}
\usepackage{dcolumn}
\usepackage{bm}
\usepackage{setspace}
\usepackage{color}

\newcommand{\Hext}{\ensuremath{H_{\mathrm{ext}}}}

\newcommand{\Hpz}{\ensuremath{H_{\mathrm{MNP}}^{\mathrm{Z}}}}

\newcommand{\Ms}{\ensuremath{M_{\mathrm{S}}}}

\begin{document}

\title{Frequency-based nanoparticle sensing over large field ranges using the ferromagnetic resonances of a magnetic nanodisc}

\author{Maximilian Albert}
\email{maximilian.albert@soton.ac.uk}
\author{Marijan Beg}
\author{Dmitri Chernyshenko}
\author{Marc-Antonio Bisotti}
\author{Rebecca L.~Carey}
\affiliation{Faculty of Engineering and the Environment, University of Southampton, Southampton SO17 1BJ, United Kingdom}
\author{Hans Fangohr}
\email{h.fangohr@soton.ac.uk}
\affiliation{Faculty of Engineering and the Environment, University of Southampton, Southampton SO17 1BJ, United Kingdom}
\author{Peter J.~Metaxas}
\email{peter.metaxas@uwa.edu.au}
\affiliation{School of Physics, M013, University of Western Australia, 35 Stirling Hwy, Crawley WA 6009, Australia}

\date{\today}

\begin{abstract}
Using finite element micromagnetic simulations, we study how resonant magnetisation dynamics in thin magnetic discs with perpendicular anisotropy are influenced by magnetostatic coupling to a magnetic nanoparticle. We identify resonant modes within the disc using direct magnetic eigenmode calculations and study how their frequencies and profiles are changed by the nanoparticle's stray magnetic field. We demonstrate that particles can generate shifts in the resonant frequency of the disc's fundamental mode which exceed resonance linewidths in recently studied spin torque oscillator devices. Importantly, it is shown that the simulated shifts can be maintained over large field ranges (here up to 1\,T). This is because the resonant dynamics (the basis of nanoparticle detection here) respond directly to the nanoparticle stray field, i.e.~detection does not rely on nanoparticle-induced changes to the magnetic ground state of the disk. A consequence of this is that in the case of small disc-particle separations, sensitivities to the particle are highly mode- and particle-position-dependent, with frequency shifts being maximised when the intense stray field localised directly beneath the particle can act on a large proportion of the disc's spins that are undergoing high amplitude precession.
\end{abstract}

\maketitle

\section{Introduction}

\begin{figure}
  \centering
  \includegraphics[width=7cm]{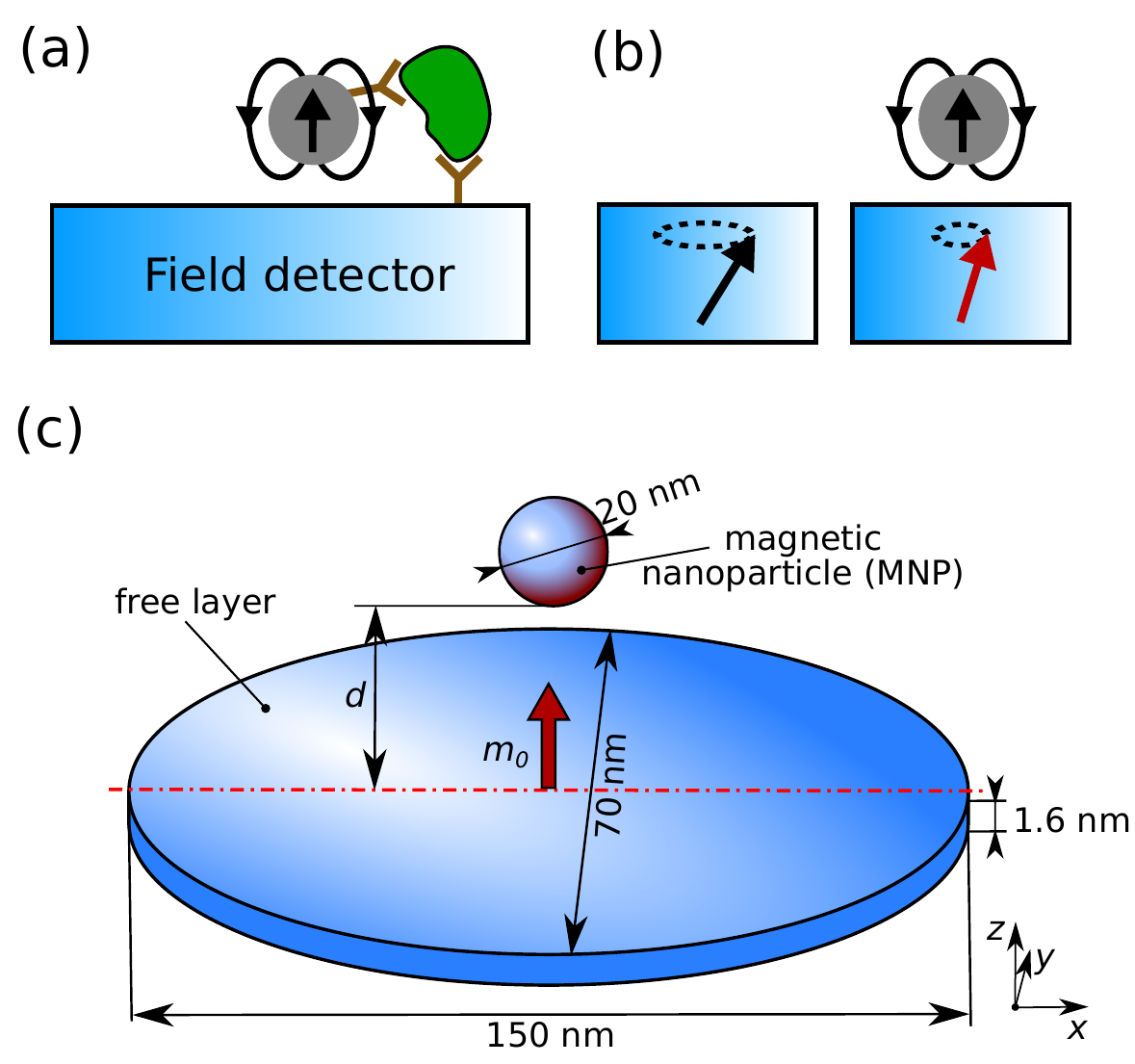}
  \caption{\label{fig:geometry} (a) Schematic of a biological entity bound to a magnetic field detection device and a magnetic nanoparticle. (b) Schematic of precessional magnetisation dynamics in a bare nano-element (left), and in the presence of a particle which changes the local field and thus the resonant dynamics (right). (c) The geometry used for the finite element simulation, composed of a magnetic nanoparticle and an elliptical magnetic disc. The red arrow indicates the equilibrium magnetisation~$\mathbf{m}_{0}$ of the disc, which points out-of-plane due to perpendicular anisotropy. The centre of the disc is located at $(x,y)=(0,0)$.}
\end{figure}

Nano-magnetic and spintronic technologies find application in various sensing scenarios~\cite{Wolf2001,Tulapurkar2005,Diegel2009}. Their appeal in \textit{biological} sensing or `biosensing' comes partly from the fact that most biomedical samples have a negligible magnetic background enabling matrix-insensitivity~\cite{Gaster2009}. This enables the use of magnetic nanoparticles (MNPs) to tag and subsequently detect biological analytes of interest reliably within a range of bodily fluids~\cite{Gaster2009,Srinivasan2011,Lee2015}.  Common device setups include sandwich assays~\cite{Gaster2009,Srinivasan2009,Krishna2016}, where the analyte is immobilised on the sensor surface and sandwiched between two antibodies which bind it to the sensor and the MNP used for detection (Fig.~\ref{fig:geometry}(a)), and flow cytometry~\cite{Helou2013,Loureiro2011a}, where the MNP-tagged analyte is detected as it flows through a microfluidic channel. While numerous techniques for particle detection exist~\cite{Chemla2000,Miller2002,Ejsing2005,Nikitin2007,Donolato2009,Chung2013,Devkota2013}, there has been a strong focus on electronic field sensing technologies employing magnetoresistive stacks~\cite{Gaster2009,Baselt1998,Osterfeld2008,Srinivasan2009,Hall2010,Martins2010,Srinivasan2011,Freitas2012,Wang2013a,Li2014,Krishna2016,Helou2013,Loureiro2011a} or Hall effect devices~\cite{Besse2002,DiMichele2011}. These devices can be used to detect small variations in magnetic field, including those generated by functionalised MNPs, converting the presence of a MNP to a change in the device resistance (typically measured as a voltage). 

In conventional magnetoresistive sensors,  MNPs induce a change in the static magnetisation configuration of the device's active sensing layer~\cite{Lee2016}. This   translates to a change in the device resistance, enabling electronic nanoparticle detection. However, one may also exploit the magnetic field dependence of ferromagnetic resonance for detection of magnetic fields~\cite{Braganca2010,Inoue2011,Atalay2015,Mizushima2010,Ryan2011} and thus magnetic nanoparticles~\cite{Ryan2011,Metaxas2015,Sushruth2016,Wohlhuter2015,Fried2016,Srimani2015,Manzin2016}. Notably, the ferromagnetic resonance frequency within the device will respond directly to  the field of the MNP  (Fig.~\ref{fig:geometry}(b)), even when the underlying magnetic ground state is unchanged. This opens pathways to intrinsically frequency-based detection schemes~\cite{Braganca2010,Mizushima2010}. Potential advantages of a frequency-based, dynamic technique over static magnetoresistive sensing include a larger  field range over which the device response is linear~\cite{Petrie2014,Rippard2010,Braganca2010} (enabling larger fields to be applied and thus generate higher MNP moments),  intrinsically frequency-based operation (typically at GHz frequencies and thus far from low frequency $1/f$ contributions), the lack of d.c.~voltage-level drift (when using direct frequency measurement) and  excellent size scalability~\cite{Braganca2010}. Electrical read-out of dynamics in isolated devices can be carried out using spin torque oscillators (STOs, where d.c.~current is used to  drive magnetisation dynamics that can then be detected in real time magnetoresistively~\cite{Krivorotov2005,Zeng2012a,Grimaldi2014}) or by using devices exploiting the inverse spin Hall effect~\cite{Saitoh2006} which enables voltage-level-based read-out of dynamics. Although the latter is not a direct frequency-based method (in that dynamics are sensed using a voltage), such a technique can still benefit from the reduced device sizes  and larger operational field ranges which come with a switch to sensing based on magnetisation dynamics.

In this work, we use micromagnetic simulations and eigenmode evaluation to quantify the effect of a MNP on resonant modes of precessional magnetisation dynamics in an underlying, out-of-plane magnetised ferromagnetic nanodisc. 
The simulations demonstrate that the MNP can induce large  shifts of the mode frequencies for the disc's quasi-uniform  and higher order resonances. At small separations, the shifts notably depend strongly on  the position of the particle relative to the regions in the disc where the dynamics associated with each particular mode are concentrated. This is a result of the non-uniform but intense field directly beneath the MNP acting on  resonant dynamics within the disc which are also highly spatially non-uniform. At larger separations however, the stray field is weaker and more uniform over the length scale of the disc. This results in a weaker dependence of the frequency shift on the lateral particle position and, as a result, similar responses for all modes (despite their different localisations within the disk). We will also demonstrate that strongly increasing the external field does not  significantly compromise device sensitivity (measured in terms of the magnitude of the MNP-induced frequency shift).

The paper is organised as follows. In Sec.~\ref{smethods} we will discuss the system and the simulation method. In Sec.~\ref{sbare} we present the first five eigenmodes of the bare disk, including spatial profiles and field dependencies of the modes in the presence of \textit{spatially uniform}, out-of-plane magnetic fields. We then present results on   MNP-modified modes (Sec.~\ref{schange}) and show how these modifications  depend on both the position of the particle and the profiles of each mode (Sec.~\ref{sposition}). Finally, we discuss the dependence of the frequency shifts on particle parameters and external field (Sec.~\ref{smaximise}).

\section{Methods}\label{smethods}

In Fig.~\ref{fig:geometry}(c) we show the simulated 1.6~nm thick elliptical nanodisc which has major and minor axis lengths of 150~nm and 70~nm, respectively.  A spherical MNP of diameter 20~nm is located above the disc. The separation between the upper surface of the disc and the bottom surface of the MNP is denoted by $d$. The nanodisc approximates the CoFeB free layer of a STO shown recently to function under low injected current and without strong external magnetic fields~\cite{Zeng2013}, properties which may be advantageous for low-power, portable diagnostics~\cite{Li2014}. Indeed exploiting precession of out-of-plane moments in STOs  offers excellent potential in terms of achieving low linewidth outputs~\cite{Rippard2010,Kubota2013,Maehara2014}, something which is critical for distinguishing MNP-induced frequency changes. Since we look only at dynamics within the disc, our results can be equally well applied to (arrays of) discs probed using inductive techniques~\cite{Metaxas2015} or, as mentioned above, via the inverse spin Hall effect~\cite{Saitoh2006} (rather than the  dynamic magnetoresistive techniques exploited in STO measurements).  

\begin{figure}[htbp]
  \centering
  \includegraphics[width=8cm]{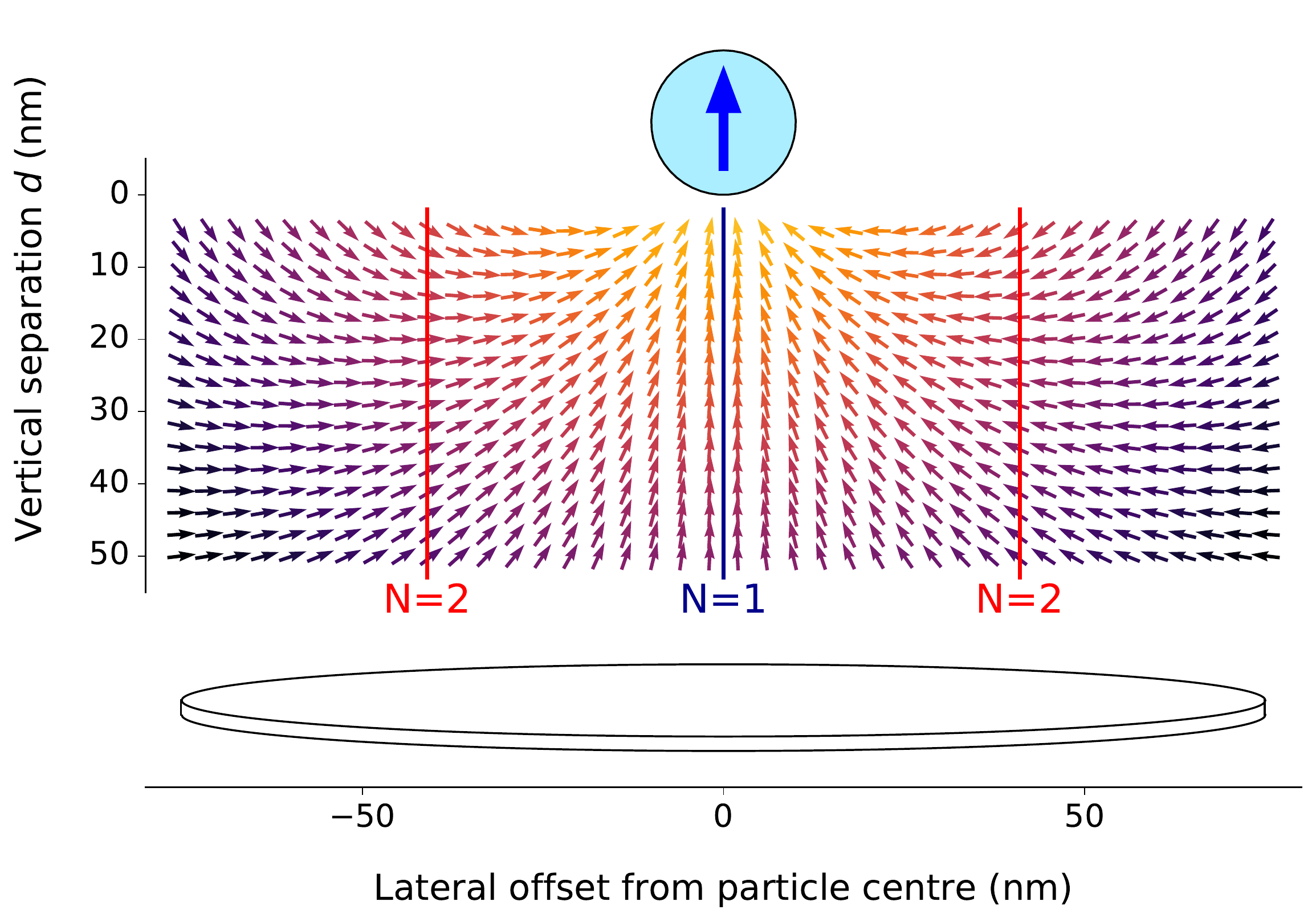}
  \caption{\label{fig:dip_field} Vector field plot of the dipole field generated by a uniformly $z$-magnetised MNP. The vectors are scaled to uniform length with their colour indicating the field strength (orange is high and violet/black is low). The vertical lines correspond to the $x$-values where modes 1 and 2 have maxima in their spin precession amplitude (see Fig.~\ref{fig:eigenmodes}(a)-(b)). A schematic of the nanodisc is shown at the bottom. 
	}
\end{figure}

We use the following simulation parameters for the nanodisc~\cite{Zeng2013}: saturation magnetisation $\Ms=1.1$~MA/m, exchange stiffness $A=20$~pJ/m and perpendicular magnetic anisotropy constant $K_1=0.74$~MJ/m$^3$. The equilibrium magnetisation $\mathbf{m}_{0}$ in the bare disc is aligned with the (out-of-plane) $+z$-direction, due to the perpendicular anisotropy (see Fig.~\ref{fig:geometry}(c)). We also model a `generic' spherical MNP with saturation magnetisation of $1$~MA/m. This magnetisation is higher than that generally expected for saturated iron oxide MNPs ($\sim 0.25$~MA/m) but lower than that observed in high moment FeCo systems~\cite{Bai2005} ($\sim 1.8$~MA/m assuming a density of $\sim 8.3$ g/cm$^3$). Its diameter will be 20 nm unless otherwise specified. The dipole field of the MNP, when uniformly $z$-magnetised, is shown in  Fig.~\ref{fig:dip_field}. The strong $z$-component of the MNP's field directly beneath the MNP is clearly visible. Unless otherwise noted, simulations have been run with an external $+z$ field of $0.1\,$T.

We use a finite element based micromagnetic simulation tool inspired by the Nmag package~\cite{Fischbacher2007} and based on the FEniCS libraries~\cite{AlnaesBlechta2015a}. Sumatra~\cite{SumatraWebpage}, IPython~\cite{PerezIPython2007}, the Jupyter notebook~\cite{JupyterWebpage}, numpy/scipy~\cite{Scipy}, pandas~\cite{McKinney2010}, matplotlib~\cite{Hunter2007} and HoloViews~\cite{Stevens2015} have been used for simulation capture and data analysis. For the computation of the resonant magnetisation modes of the system we employ an eigenvalue problem-based method  used recently in Refs.~\onlinecite{Metaxas2015,Metaxas2016} which is similar to that presented by d'Aquino \textit{et al.}~\cite{dAquino2009}. Firstly, we compute the system's equilibrium configuration $\mathbf{m}_{0}(\mathbf{r})$ and then linearise the Landau-Lifshitz equation for magnetisation dynamics around $\mathbf{m}_{0}(\mathbf{r})$. This results in a system of linear differential equations for resonant oscillations of the magnetisation, $\mathbf{dm}(\mathbf{r},t)$, occurring around $\mathbf{m}_{0}(\mathbf{r})$. This system of differential equations can be written as an eigenvalue problem~\cite{Metaxas2015}. The eigenvectors correspond to the resonant eigenmodes~\cite{Baker2016} of the nanodisc, each occurring at a resonant frequency given by the mode's eigenvalue.

The raw data for the relevant figures in this paper, as well as Jupyter notebooks~\cite{JupyterWebpage} to reproduce them from this data, are available in the associated electronic supplementary material for this paper~\cite{GithubrepoSupplementaryMaterial2016}.

\section{Results and discussion}

\subsection{Magnetic eigenmodes of a bare disc}\label{sbare}

\begin{figure}
  \centering
  \includegraphics[width=7cm]{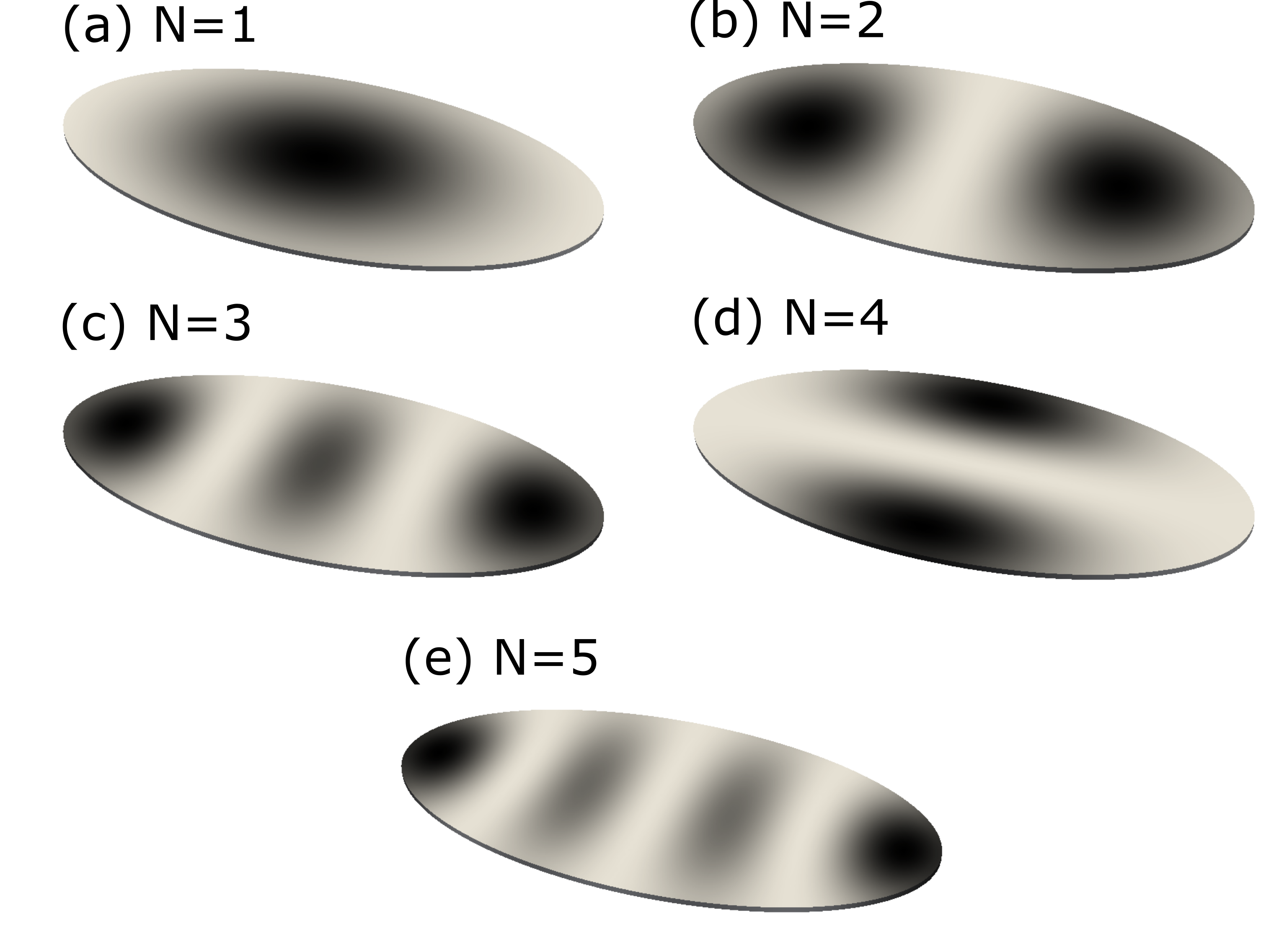}
  \caption{Profiles of the first five modes of the bare disc. Dark regions correspond to large-amplitude precessions of the dynamic magnetisation.}
  \label{fig:eigenmodes}
\end{figure}

We firstly compute the eigenmodes of the STO's elliptical free layer in the absence of a MNP. In the absence of an external field the obtained resonance frequencies of the first five modes (in order of frequency) are $f_1 =1.60\,$GHz, $f_2 =3.09\,$GHz, $f_3=5.09\,$GHz, $f_4=5.55\,$GHz and $f_5 = 7.77\,$GHz. $f_1$ is of the same order as the excited mode  reported by Zeng \textit{et al.}~\cite{Zeng2013} when extrapolating their data to the zero-current case. Representations of the spatial profiles of mode~1 (most relevant for STOs) as well as the next four higher order modes are shown in Fig.~\ref{fig:eigenmodes}. The shading encodes the amplitude of the magnetisation precession for each mode, with dark regions representing high amplitude resonant oscillations.   

The fundamental  $N=1$ mode (Fig.~\ref{fig:eigenmodes}(a)) consists of an in-phase precession of all magnetic moments. The precession amplitude is largest at the sample's centre and decays in amplitude towards the boundaries of the free layer. The $N=2$ mode (Fig.~\ref{fig:eigenmodes}(b)) is characterised by two regions of large precession amplitude in the left/right halves of the disc, with the magnetic moments in the two parts precessing out-of-phase.
Modes 3 and 5 (Figs.~\ref{fig:eigenmodes}(c,\,e)) both have multiple nodal axes parallel to the short axis of the ellipse. Mode 4 (Fig.~\ref{fig:eigenmodes}(d)) is similar to mode 2 but with a nodal axis along the long axis of the disc. 

\begin{figure}
  \centering
  \includegraphics[width=7cm]{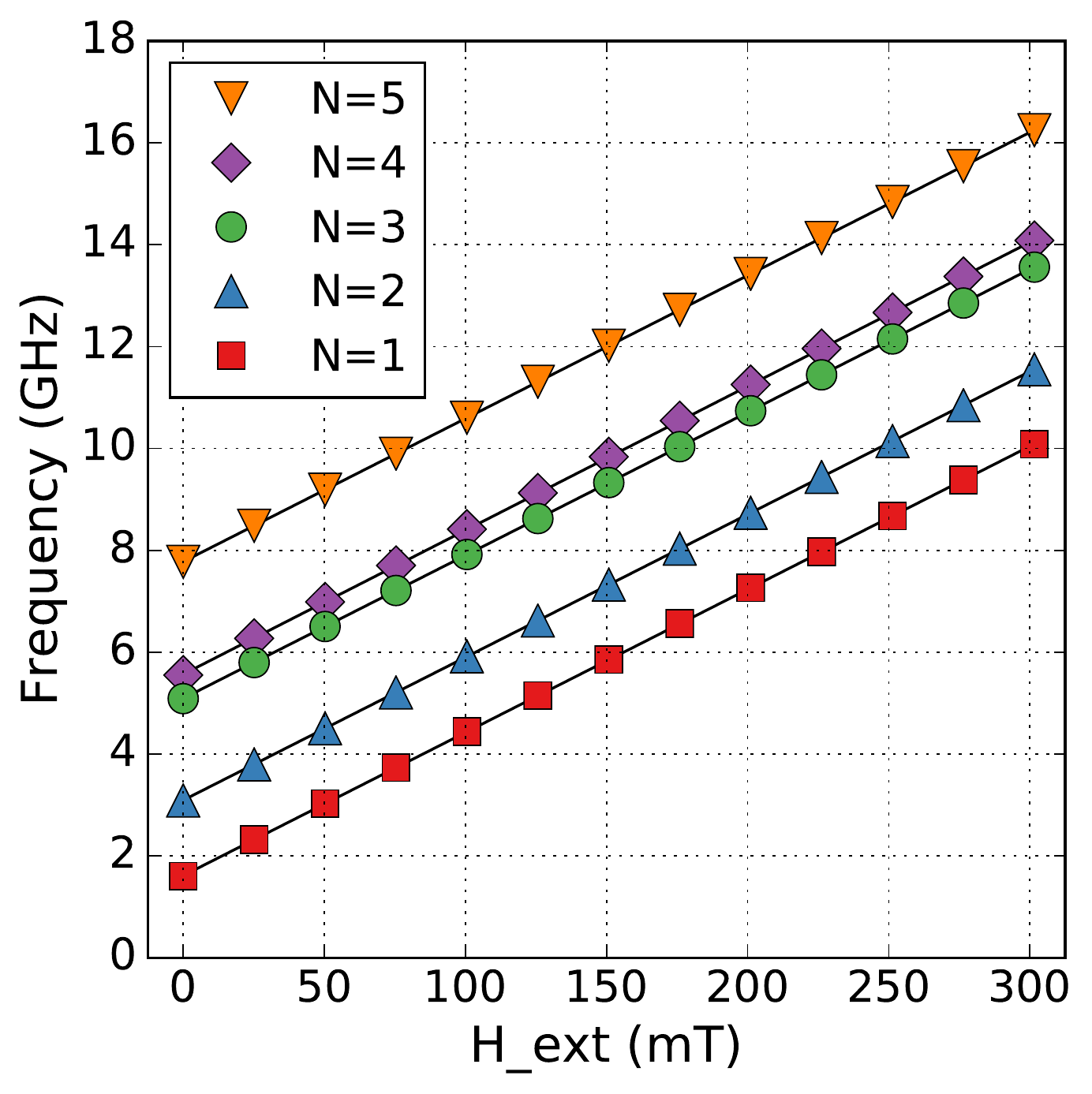}
  \caption{\label{fig:Hdep} Out-of-plane ($+z$) field dependence of the frequencies of the first five modes of the bare disc. Linear regressions (black lines) fitted to the data points demonstrate that the field-induced changes in each eigenfrequency are equivalent to one another to within $0.8$\%.}
\end{figure}

In Fig.~\ref{fig:Hdep} we show the dependence of each mode frequency on the magnitude of a spatially uniform magnetic field applied along $+z$ (aligned with $\mathbf{m}_{0}(\mathbf{r})$). The extracted field sensitivities (frequency shift per unit field) of all modes are consistent with one another to within $0.8$\% and have a value of $28.162 \pm 0.108$~MHz/mT. As will be shown below, however, the situation is more complex in the presence of a small MNP due to the combination of the  localised  MNP stray field and spatially non-uniform mode profiles. This will generate a clear dependence of the frequency shifts on the position of the MNP relative to the disc (both in the lateral and vertical directions). 

\subsection{MNP-modified eigenmodes}\label{schange}

We now introduce a magnetic nanoparticle to the system as per Fig.~\ref{fig:geometry}(c). It is magnetised in the $+z$-direction, parallel to the applied external field. 

\begin{figure}
  \centering
  \includegraphics{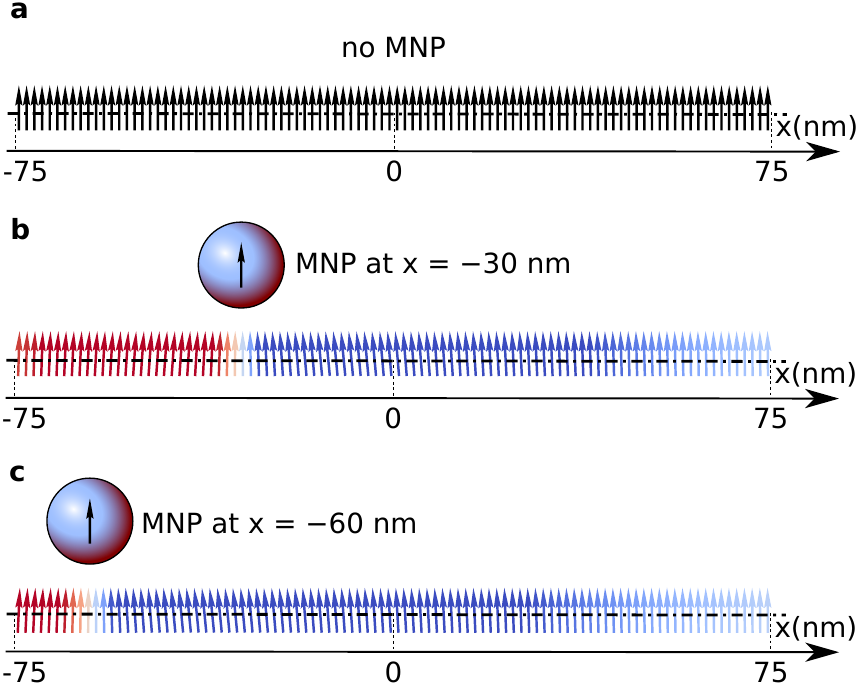}
  \caption{\label{fig:cross_section} Cross section of the equilibrium magnetisation configuration along the $x$-axis in absence of a MNP (a), and in presence of a MNP ($d=5$~nm) above the major axis of the disc, at $x=-30$~nm (b) and $x=-60$~nm (c). The colours represent the amount of canting of the magnetisation (magnitude and direction of the $m_x$-component).}
\end{figure}

In the equilibrium configuration of the bare disc, the disc's free layer magnetisation is parallel to the (out-of-plane) $z$-axis everywhere due to the perpendicular anisotropy. This is illustrated in Fig.~\ref{fig:cross_section}(a), which shows the magnetisation vector field along the disc's long $x$-axis. If a MNP is present however, its stray field locally modifies the magnetisation. Examples of this are shown in Figs.~\ref{fig:cross_section}(b) and \ref{fig:cross_section}(c) where the MNP is located at $x = -30 \,\text{nm}$ and $x = -60 \,\text{nm}$, respectively, above the major axis  ($y = 0$) of the disc ($d=5$~nm) The presence of the localised MNP field  results in a slight localised canting of the magnetisation towards the MNP due to the in-plane components of the stray field. Note, however, that the actual change in the magnetisation orientation is very small. For example, at separation $d=5$~nm the spatially averaged, normalised out-of-plane magnetisation is $0.99874$ and at $d=20$~nm it is $0.99981$. These are both very close to $1$, which is the value expected for a perfectly out-of-plane magnetised disc. However, despite only inducing a very small change in the magnetisation configuration, the stray field from the MNP can generate strong changes in the resonant frequency (up to $\sim 350$ MHz for $N=1$ at $d=5$~nm), as will be shown below. 

\begin{figure}
  \centering
  \includegraphics[width=7cm]{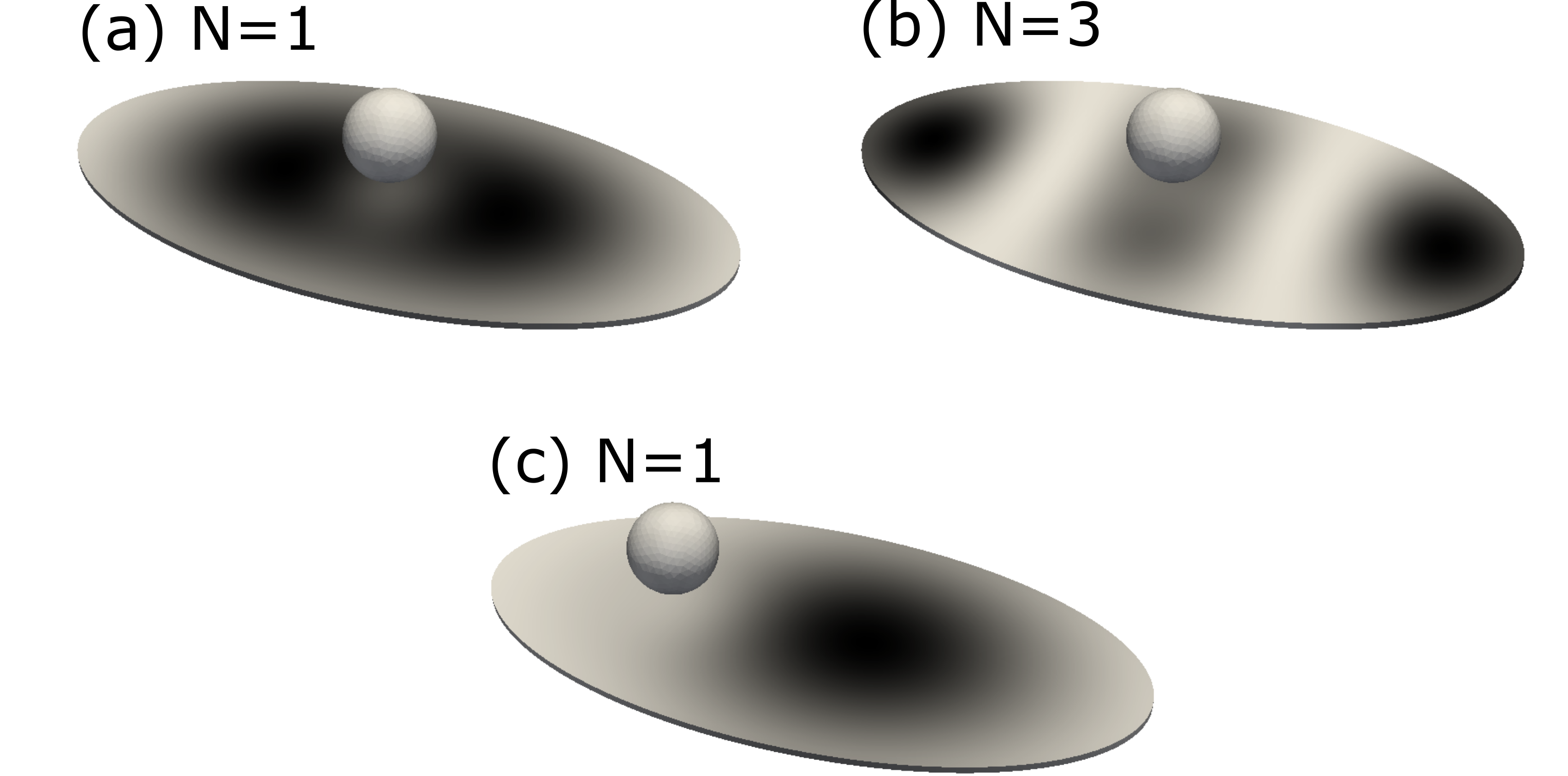}
  \caption{Mode profiles with a MNP centred above the disc for $N=1$ (a) and $N=3$ (b) as well as for an off-centre MNP for $N=1$ (c). The MNP-disc separation, $d$, is 5~nm. Dark regions correspond to large-amplitude precessions of the dynamic magnetisation.
	}\label{fig:eigenmodes2} 
\end{figure}

Beyond modifying the resonance precession frequency of the magnetisation, the non-uniformity of the MNP field can also noticeably modify  the spatial profile of the eigenmode. An example for the $N=1$ mode is shown in Fig.~\ref{fig:eigenmodes2}(a), with a MNP at $d=5$~nm. Although the mode excitations remain in-phase over the entire disc there is a reduced oscillation amplitude directly beneath the MNP, consistent with a MNP-induced, local stiffening of the magnetic moment.  This also occurs for the $N=3$ mode (Fig.~\ref{fig:eigenmodes2}(b)), which is the only other  mode studied here that has an antinode (= location of maximum oscillation amplitude) at the disc's centre. Similarly, a left displaced particle (Fig.~\ref{fig:eigenmodes2}(c)) will lead to larger oscillation amplitudes for the $N=1$ mode on the opposing (right) side of the disc.
Somewhat analogously for vortices, a localised out-of-plane field at the core of a vortex can stiffen the core, increasing the frequency of its gyrotropic mode in the small displacement limit~\cite{Fried2016}.

\subsection{MNP position and height dependence}\label{sposition}

We now look more closely at the influence of the position of the MNP on the frequency shifts, $\Delta f$. In the case of small MNP-disc separations ($d \lessapprox 15$~nm), there is an intense, localised and  predominantly $+z$ oriented field directly beneath the MNP (Fig.~\ref{fig:dip_field}). As a result, at these small separations
the effect of the MNP on a mode is determined by where the mode's dynamics are concentrated relative to the MNP's location. For MNPs directly above regions where there are high precession amplitudes, there is a frequency upshift ($\Delta f > 0$) due to the increased local $z$-field below the disc (as per Fig.~\ref{fig:dip_field}). This result is qualitatively consistent with that seen for uniform fields in Fig.~\ref{fig:Hdep}. However, when dynamics are occurring in a region which is laterally offset from the particle and the particle is very close to the disc, the precessing moments can be subject to a field oriented in the $-z$--direction (see again Fig.~\ref{fig:dip_field}), leading to a \textit{reduction} in the frequency ($\Delta f < 0$).  We discuss these behaviours below.

\begin{figure*}
  \centering
  \includegraphics[width=14cm]{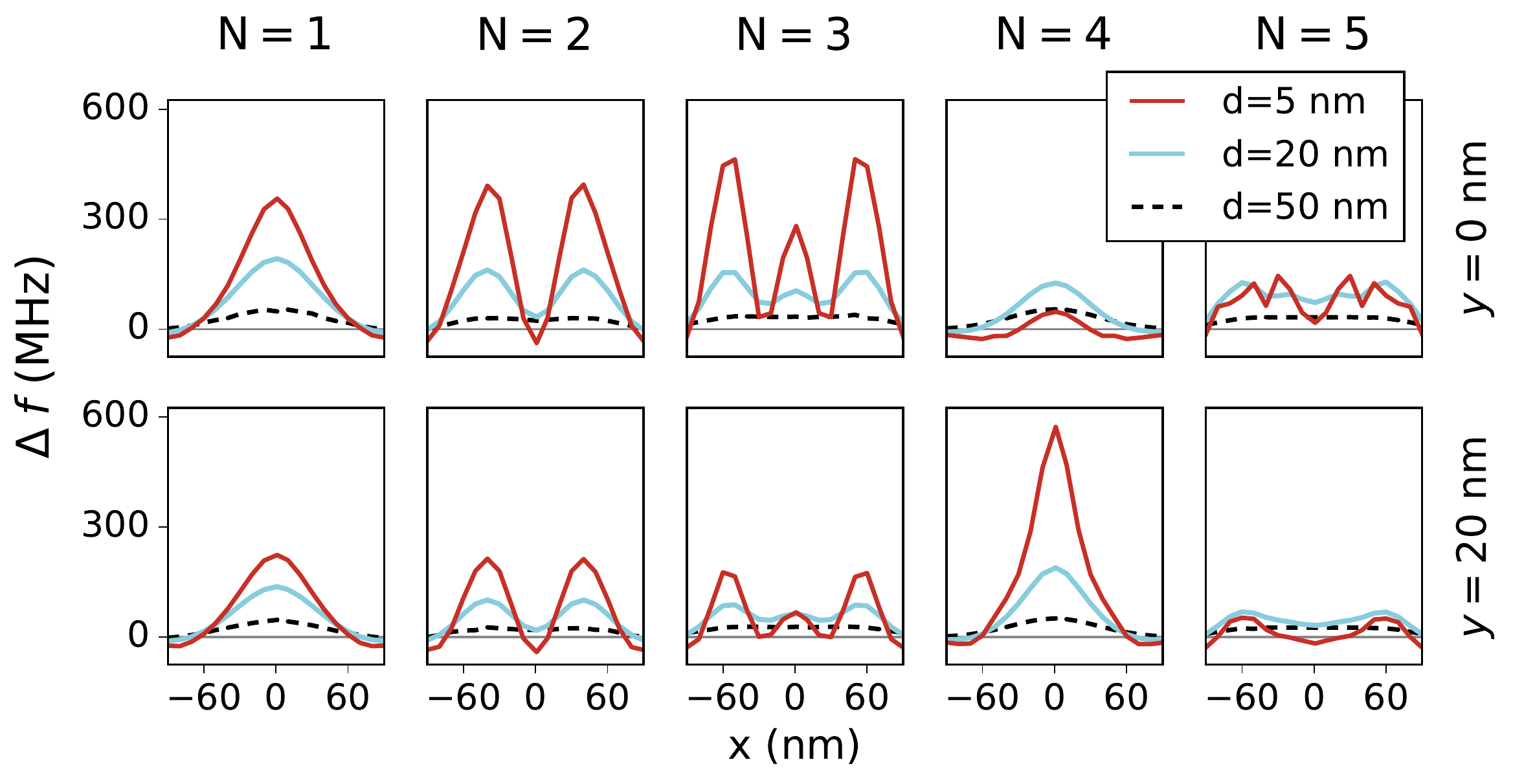}
  \caption{Frequency change $\Delta f$ as a function of lateral particle position $x$ for the first five eigenmodes ($N=1-5$), with the MNP either above the major axis of the ellipse (top row) or shifted by 20~nm in $y$-direction (bottom row). The frequency change for each mode is shown for three values of the MNP-disc separation (red: $d=5$~nm, blue: $d=20$~nm, dashed black: $d=50$~nm).}
  \label{fig:sensitivities_on_lateral_MNP_pos}
\end{figure*}

Fig.~\ref{fig:sensitivities_on_lateral_MNP_pos} shows the $\Delta f$ values observed for all five eigenmodes ($N=1,\ldots,5$) when shifting the MNP along the disc's long axis at separations of $d=5$, 20 and 50~nm, for lateral MNP positions of $y=0$~nm and $y=20$~nm. If the MNP is close to the surface of the disc ($d=5$~nm, red lines in Fig.~\ref{fig:sensitivities_on_lateral_MNP_pos}) then  $\Delta f$  closely follows the spatial mode pattern, as can be seen by comparing the red lines in Fig.~\ref{fig:sensitivities_on_lateral_MNP_pos} with Fig.~\ref{fig:eigenmodes}(a)-(e). This reflects the fact that in this case the MNP stray field has a very localised influence on the underlying precessing moments. For example, for $N=1$ there is a single, broad sensitivity peak near the centre of the disc ($x=0$~nm) for both values of $y$, which mirrors that mode's spatial profile. Likewise, the $N=2$ mode exhibits two sensitivity peaks near $x\approx\pm 40$~nm, i.e.~near the locations of the mode's antinodes (see Fig.~\ref{fig:eigenmodes}(b)). The curve for $N=3$ shows three such peaks, with the outer ones  slightly higher, reflecting the fact that the outer antinodes of this mode have a larger amplitude than the middle one (see Fig.~\ref{fig:eigenmodes}(c)). 

This pattern continues for $N=4$ and $N=5$, but while the modes $N=1,2,3$ show fairly similar sensitivities for $y=0$~nm and $y=20$~nm, those for $N=4,5$ are quite different for both values of~$y$, reflecting the fact that these two modes are less uniform along the short axis of the disc. This is very obvious for $N=4$ where the mode antinodes are located at $y\approx\pm 25$~nm (Fig.~\ref{fig:eigenmodes}(d)), leading to a large positive $\Delta f$ for the laterally $y$-offset MNP (bottom plot for $N=4$ in Fig.~\ref{fig:sensitivities_on_lateral_MNP_pos}). 

Comparing the curves obtained for different $d$ values in Fig.~\ref{fig:sensitivities_on_lateral_MNP_pos}, it is evident that with increasing disc-MNP separations, the shifts are reduced and the curves become more homogeneous and indeed comparable for each mode. For an intermediate separation of $d=20$~nm, the influence of the mode profiles on the MNP-position-dependent frequency shifts are still detectable. However, at a larger separation of $d=50$~nm, the $\Delta f$ curves for each mode become very similar (those for $N=1,4$ show a flat, broad peak around the centre of the disc whereas those for $N=2,3,5$ have a more plateau-shaped profile, but the difference is small). This is consistent with two factors: (i) a much more uniform MNP field across the disc at large $d$ (Fig.~\ref{fig:dip_field}) (with a net $+z$-orientation leading  to a frequency increase); and (ii) almost identical sensitivities for each mode in a uniform field (Fig.~\ref{fig:Hdep}), leading to similar $\Delta f$ values for each mode in the presence of the more uniform MNP field.
We note that in the presence of multiple MNPs at similar separations (as they might occur in real devices) this will likely result in a $\Delta f$ that is (roughly) proportional to the number of nanoparticles present, due to the fact that all of them induce a similar frequency change, independent of their exact location above the disc.

\begin{figure}
  \centering
  \includegraphics[width=8cm]{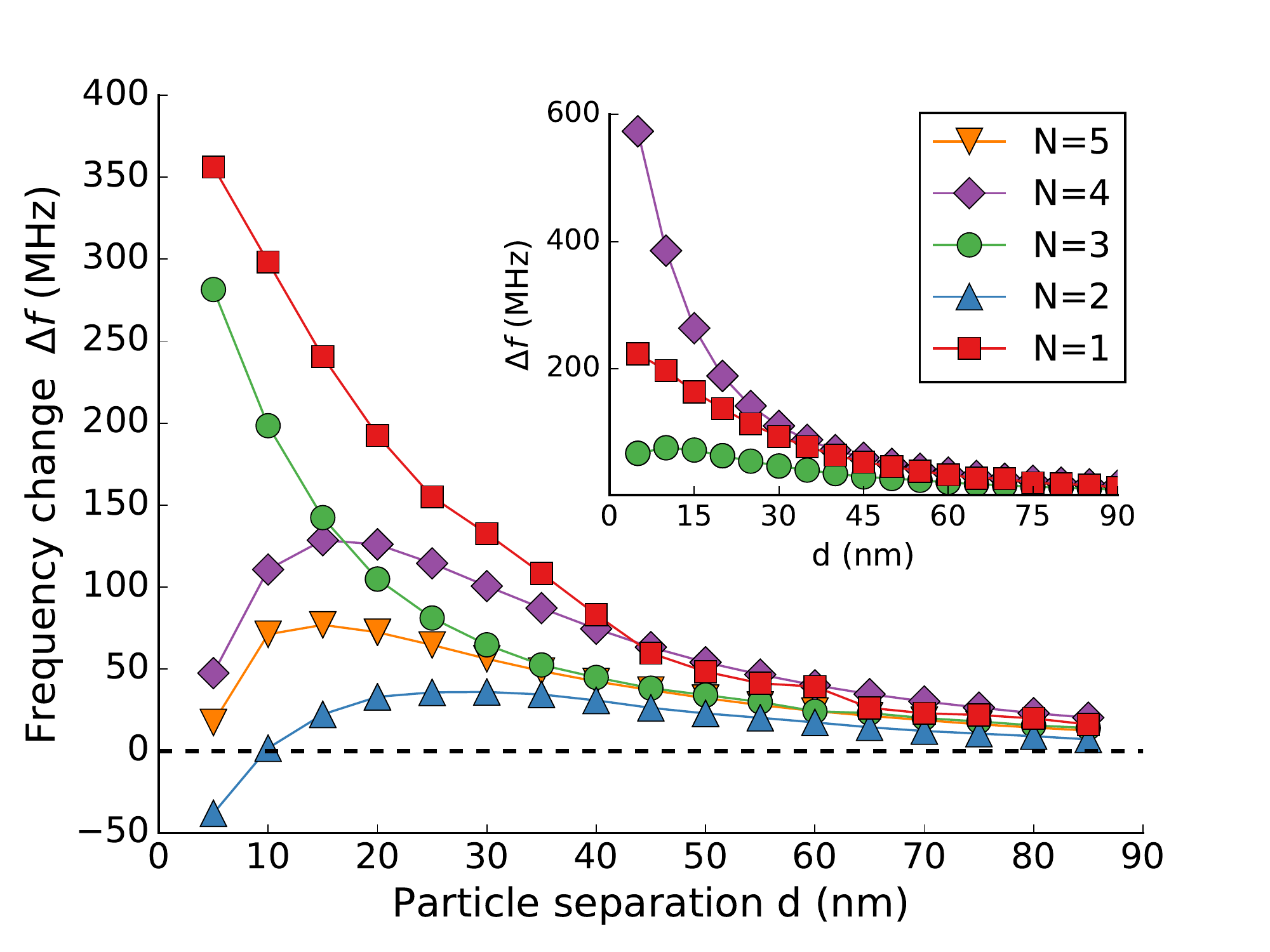}
  \caption{Frequency change $\Delta f$ as a function of particle separation $d$ for a MNP located above the centre of the free layer at $(x,y)=(0, 0)$. The inset shows $\Delta f$ for the first three modes as a function of $d$ for an off-centre particle at $(x,y)=(0, 20\,\textrm{nm})$.}
  \label{fig:distdep}
\end{figure}

Fig.~\ref{fig:distdep} shows how $\Delta f$ varies as the separation~$d$ is varied for a laterally centred MNP (i.e., one which is laterally positioned at $(x,y)=(0,0)$). Consistent with the $d=5$~nm data in Fig.~\ref{fig:sensitivities_on_lateral_MNP_pos}, the $N=1$ and $N=3$ modes, which both have dynamics concentrated below the centred MNP, exhibit the largest $\Delta f$ at small separations. Furthermore, due to the concentration of dynamics below the MNP, the shifts are positive for these two modes for all $d$ values. This is because as $d$ is varied there is no sign change in the out-of-plane component of the MNP stray field, $\Hpz$, acting on the central precessing moments in the disc (cf. blue vertical line in Fig.~\ref{fig:dip_field}).  At large distances, all other modes also exhibit a positive $\Delta f$ which decreases with increasing $d$, again consistent with what is seen in Fig.~\ref{fig:sensitivities_on_lateral_MNP_pos} and discussed above.  Note that although the mode $N=3$ also has an antinode at the disc centre, its $\Delta f$   is smaller than that seen for N=1 because the the N=3 dynamics are distributed amongst three antinodes (cf. Fig.~\ref{fig:eigenmodes}(c)), with the outer antinodes being less strongly affected by the centralised MNP.

One can also see  in Fig.~\ref{fig:distdep} that the frequency shift for modes 2, 4 and 5 \textit{decreases} as $d$ approaches zero. This is because these modes have dynamics concentrated away from the centre of the disc and they are thus exposed to a weaker or even a negative $\Hpz$ at small MNP-disc separations. For example, the frequency shift for $N=2$ is zero at $d\approx 10$~nm (blue triangles in Fig.~\ref{fig:distdep}). At this value of $d$, $\Hpz$ is indeed $\approx 0$ near the location where the mode dynamics are concentrated ($x\approx \pm 41$~nm; see the vertical red lines in Fig.~\ref{fig:dip_field}). Note that the exact value of $|x|$ where $\Hpz$ is zero is slightly smaller than 41~nm. This is however not unexpected because the frequency shift will result from what is effectively a convolution between the mode profile and the particle stray field whose magnitude is non-uniform across the disc. We also note that at  $x\approx \pm 41$~nm $\Hpz$ becomes negative for values of $d$ smaller than $\approx 15$~nm  (Fig.~\ref{fig:dip_field}). This is consistent with the observed negative frequency shift for the N=2 mode at very small $d$ in Fig.~\ref{fig:distdep}. Regarding the $N=4,5$ modes, because they have antinodes located closer to the lateral centre of the disc than the $N=2$ mode (Fig.~\ref{fig:eigenmodes2}) they do not experience a null $\Hpz$ until even smaller MNP-disc separations (Fig.~\ref{fig:dip_field}). This means that they have a $\Delta f=0$ crossing in Fig.~\ref{fig:distdep} at smaller $d$. Finally, we note that for a particle shifted in the $y$-direction (and thus lying above the antinode  of the $N=4$ mode), it is the $N=4$ mode which exhibits the highest $\Delta f$  (see purple diamonds in inset of Fig.~\ref{fig:distdep}).

\subsection{System dependencies of $\Delta f$}\label{smaximise}

For sufficiently large read-out signals, detection based on identifying changes to the resonant frequency of a device will ultimately be limited by the resonance linewidth as well as the MNP-induced $\Delta f$, with a small linewidth and large $\Delta f$ being optimal. In the STO study of Zeng \textit{et al.}~\cite{Zeng2013}, the minimum observed linewidth of the primary mode (corresponding to the $N=1$ mode here)  was on the order of 30 MHz, suggesting that detection could be realised for $d$ up to $\sim 50$ or $60$~nm based on the $\Delta f$ values shown in Fig.~\ref{fig:distdep}. This assumes however that such linewidths can be maintained under the fields required to magnetise the MNP and that passivation layers and/or upper contacts can be made sufficiently thin (this is reasonable given that typical passivation layers are on the order of 30-50 nm thick~\cite{Krishna2016,Lee2015} and coating layers for MNP biofunctionalisation can be made very thin, on the order of 2-5~nm~\cite{Lee2015,Rafati2013}). We note that lower linewidths~\cite{Maehara2014} (e.g.~6 MHz full-width-half-maximum~\cite{Rippard2010}) have been observed in out-of-plane magnetised STOs. We also note that  oscillators based on magnetic vortices can offer even lower linewidths~\cite{Pribiag2007,Locatelli2011,Lebrun2014}, but they also typically have lower field sensitivities, highlighting the need to optimise the sensitivity-to-linewidth ratio if sensing is to be done by directly identifying changes to the frequency. The sensitivity of the specific resonance of the device to localised fields~\cite{Fried2016} is also critical of course.

\begin{figure*}
  \centering
  \includegraphics[width=17cm]{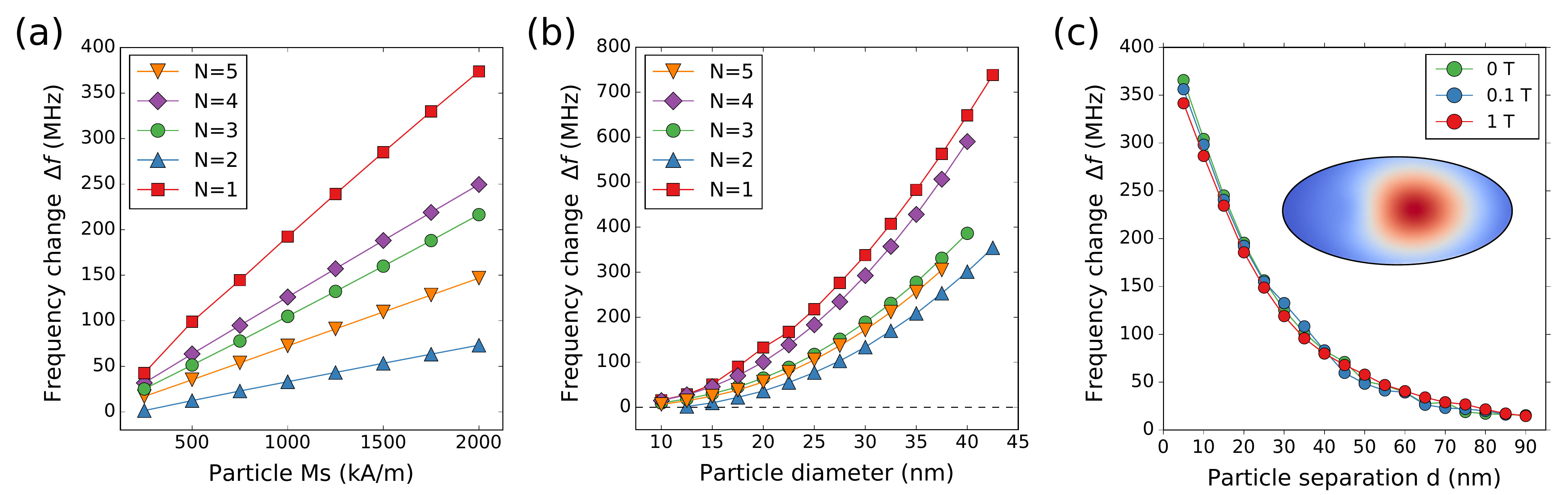}%
  \caption{%
    (a)~Dependence of $\Delta f$ on particle $\Ms$ for the first five modes, with a particle located above the centre of the disc at separation $d=20$~nm.
    (b)~Dependence of $\Delta f$ on particle size for the first five modes, with the separation between the bottom of the particle and the disc surface held constant at $d=30$~nm.
    (c)~$\Delta f$ for $N=1$ for different out-of-plane fields (other simulation parameters unchanged from Fig.~\ref{fig:distdep}). Inset shows the profile of the fundamental mode $N=1$ in an external field of strength $\mu_0 H=1$~T with a particle (not shown) off-centre at $x=-30$~nm, $d=5$~nm.
  }
  \label{fig:Msdep}
\end{figure*}

There are a number of ways to increase $\Delta f$ without modifying the properties of the nanodisc that is being used as a detector.
For example, one can  attempt to engineer particles with higher moments (see Fig.~\ref{fig:Msdep}(a), which shows $\Delta f$ versus particle moment) or increase the size of the particles (see Fig.~\ref{fig:Msdep}(b), showing $\Delta f$ versus particle diameter). In both cases, this increases the MNP-generated magnetic stray fields and thus the resultant shifts. However, oft-used iron-oxide particles will have lower moments and thus generate lower shifts. We note that in contrast to the case of magnetic vortices~\cite{Fried2016}, for this system we saw monotonic increases in $\Delta f$ when increasing the particle size (and moment). 

The time-averaged moment of superparamagnetic particles can also be increased by  increasing the applied field. Thanks to the continued linearity of  $\Delta f$ as a function of the external field strength $\Hext$ over a large range of field values (Fig.~\ref{fig:Hdep}), the external field can be increased without significantly compromising $\Delta f$. Indeed, we see good consistency between the calculated $\Delta f$ values obtained at vastly differing fields of $0\,$T, $0.1\,$T and $1\,$T (shown for $N=1$ in Fig.~\ref{fig:Msdep}(c)). Note that here we (unphysically) assume the same particle moment at each field to enable direct comparison of the $\Delta f$ values. We note that at high field (1\,T), the particle still clearly modifies the distribution of the dynamics of the $N=1$ mode (shown for a laterally offset particle in the inset of Fig.~\ref{fig:Msdep}(c)).
The continued sensitivity of this system even in large external fields is in contrast to magnetoresistive sensors whose sensitivities will be almost nil when the magnetisation is \mbox{(quasi-)}uniform in a (sufficiently) high field.

\section{Conclusion}

Using finite element micromagnetic simulations, we have shown how localised magnetic fields generated by magnetic nanoparticles (typically having diameters of 20~nm) modify the spatial profiles and frequencies of confined ferromagnetic resonances in underlying out-of-plane-magnetised ferromagnetic nanodiscs. By electrically detecting these resonances, nanoparticle-induced modifications to the resonances can be exploited to create nano-scale, frequency-based nanoparticle detectors for applications such as solid-state bio-detection~\cite{Braganca2010}.

Due to the non-uniform spatial profiles of resonant mode dynamics, the observed shifts (exceeding $300$~MHz in some cases) can depend strongly on the position of the nanoparticle. This is most obvious for small disc-particle separations, where small regions of the disc will be subject to the intense magnetic field localised directly beneath the particle. In this case, the shifts are maximised when the particle is above those regions where the spins are undergoing the highest amplitude precessional dynamics. At larger separations, the disc will be subject to a weaker but more uniform field, which leads to shifts that are smaller ($\sim 20$~MHz at an 80 nm vertical separation) but also less dependent on the lateral particle position.  It was also shown that it is possible to maintain large nanoparticle-induced frequency shifts over a wide range of external fields, exploiting the fact that detection is dependent on the action of the field on the resonant dynamics rather than a change to the static magnetisation configuration within the device. The ability to detect frequency changes experimentally will depend on the linewidth of the measured signal relative to the nanoparticle-induced frequency changes. The latter can be optimised by having small particle-disc separations and/or large particle moments (with the moment being maximised when the external field is large, for optimised nanoparticle compositions, and/or of course for larger particles).

\section*{Acknowledgements}

This work was  supported by the EPSRC's Doctoral Training Centre (DTC) grant EP/G03690X/1, the Australian Research Council's Discovery Early Career Researcher Award scheme (DE120100155), the United States Air Force (Asian Office of Aerospace Research and Development, AOARD) and the University of Western Australia (Research Collaboration Award and Early Career Researcher Fellowship Support scheme). We also acknowledge the use of the IRIDIS High Performance Computing Facility, and associated support services at the University of Southampton, in the completion of this work. The authors thank Mikhail Kostylev for useful discussions.

\bibliography{refs}

\end{document}